# The magnetic origin of the metal-insulator transition in $V_2O_3$: Mott meets Slater

J. Trastoy[1,†,*], A. Camjayi[2], J. del Valle[1], Y. Kalcheim[1], J.-P. Crocombette[3], J.E. Villegas[4], M. Rozenberg[1,5], D. Ravelosona[6], and Ivan K. Schuller[1]

[1] *Department of Physics and Center for Advance Nanoscience, University of California San Diego, La Jolla, California 92093, USA*

[2] *Departamento de Física, FCEyN, UBA and IFIBA, Conicet, Pabellón 1, Ciudad Universitaria, 1428 CABA, Argentina*

[3] *CEA, DEN, Service de Recherches de Métallurgie Physique, Université Paris-Saclay, F-91191 Gif-sur-Yvette, France*

[4] *Unité Mixte de Physique, CNRS, Thales, Univ. Paris-Sud, Université Paris Saclay, 91767 Palaiseau, France*

[5] *Laboratoire de Physique des Solides, CNRS, Univ. Paris-Sud, Université Paris-Saclay, 91405 Orsay Cedex, France*

[6] *Centre de Nanosciences et de Nanotechnologies, CNRS, Université Paris-Sud, Orsay 91405, France*

[†] *Current address: Unité Mixte de Physique, CNRS, Thales, Univ. Paris-Sud, Université Paris Saclay, 91767 Palaiseau, France*

[*] *Corresponding author: jtrastoy@physics.ucsd.edu*

**Abstract:**

Despite decades of experimental and theoretical efforts, the origin of metal-insulator transitions (MIT) in strongly-correlated materials is one of the main longstanding problems in condensed-matter physics. An archetypal example is $V_2O_3$, where electronic, structural and magnetic phase transitions occur simultaneously. This remarkable concomitance makes the understanding of the origin of the MIT a challenge due to the many degrees of freedom at play. In this work, we demonstrate that magnetism plays the key dominant role. By acting on the magnetic degree of freedom, we reveal an anomalous behaviour of the magnetoresistance of $V_2O_3$, which provides strong evidence that the origin of the MIT in $V_2O_3$ is the opening of an antiferromagnetic gap in the presence of strong electronic correlations.

Electronic correlations are responsible for a multitude of remarkable phenomena in condensed matter physics[1–7]. Among the most studied phenomena is the metal-insulator transition (MIT)[2,8–12], where a material that should be metallic at all temperatures according to a non-interacting band diagram, becomes insulating in a first-order transition with a resistance change of several orders of magnitude. Understanding the MIT has been one of the outstanding challenges in condensed-matter physics in the last decades. A number of theories were advanced amongst which three stand out. Mott proposed[2] that Coulomb repulsion between electrons could split the conduction band in two by opening a charge gap. Slater[13] argued that the gap could be opened due to antiferromagnetic ordering where electronic energies become spin-dependent. Peierls[14] suggested that structural changes, such as dimerization, could alter the periodicity of the lattice giving rise to a gap.

Disentangling these three main mechanisms can be a complicated task. No other material reflects this as, vanadium sesquioxide ($V_2O_3$)[11,15,16], which has become a classic problem of strongly correlated physics. Three transitions occur coincidentally involving the three different aspects suggested by Mott, Slater and Peierls: an electronic, a magnetic and a structural transition, respectively. More specifically, below 160K the material undergoes a transition from a paramagnetic metal with rhombohedral structure, to an antiferromagnetic (AF) insulator with a monoclinic unit cell. To complicate matters further, neutron data[17] showed that AF fluctuations above the transition are observed at a wave vector different from the one present in the low-$T$ ordered state, which was interpreted as evidence of orbital order. After almost 50 years from the discovery of the transition, it remains a matter of debate what mechanism is responsible for it.

To disentangle this problem, one has to identify the relevant degrees of freedom that play a singular role at the transition. In the present study, we shall argue that the electronic gap opening in $V_2O_3$ is driven by antiferromagnetic interactions which arise from Coulomb correlations. This is revealed by an anomalous magnetoresistance across the MIT, which shows a sign change and a dramatic enhancement near the transition. We further find that these results systematically track the MIT, which we displace in temperature through controlled ion irradiation.

All these puzzling observations are captured by quantum Monte Carlo calculations of a Hubbard model within dynamical mean-field theory (DMFT)[18], which also provide a transparent physical interpretation of the experiments. Thus, our study yields clear evidence that the gap opening in $V_2O_3$ should be considered a Mott-Slater transition, shedding new light on this longstanding problem.

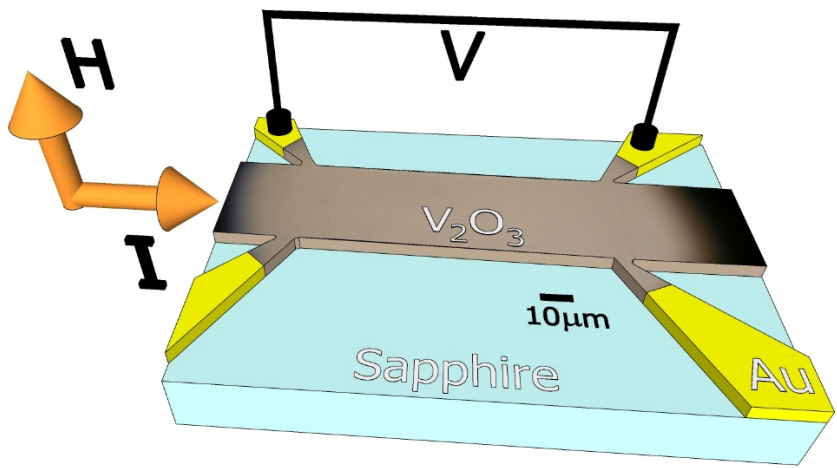

**Fig. 1 Measurement device and configuration.**

Optical microscope image of the $V_2O_3$ microbridge rendered in a 3D model. For the magnetoresistance measurements the field is applied out of plane and perpendicular to the current. The voltage is measured between the longitudinal contacts.

$V_2O_3$ microbridges (shown in Fig. 1) were fabricated from 100 nm-thick thin films through a series of photolithography and etching steps, and were subject to controlled disorder by $He^+$ ion irradiation (see methods). The temperature dependent resistivity for a pristine $V_2O_3$ bridge is shown by the black curve in Fig. 2(a). A MIT occurs around 150K as revealed by a resistivity change of 6 orders of magnitude. We explored the sensitivity of the electronic transport across the MIT to the magnetic degree of freedom. We thus measured magnetoresistance (MR) in the same pristine sample (magnetic field perpendicular to the plane, see Fig. 1), shown as the black line in Fig. 2(b) and its inset. Qualitatively similar results were obtained for in-plane field measurements parallel and perpendicular to the current $\vec{I}$ (not shown). The inset shows a series of resistivity $\rho$ vs field $H$ curves at different temperatures. The field dependence is quadratic. As expected, the MR is positive at room temperature but, surprisingly, it decreases and changes signs with decreasing temperature. In the main panel of Fig. 2(b), the black curve shows the temperature dependence of the MR at 9T, $[\rho(9T) - \rho(0)]/\rho(0)$ for the same pristine sample. The MR decreases without saturation, and seemingly diverges as the transition takes place. It is important to note that no MR measurements could be performed at lower temperatures, that is in the highly insulating state, due to the noise induced by the high resistance.

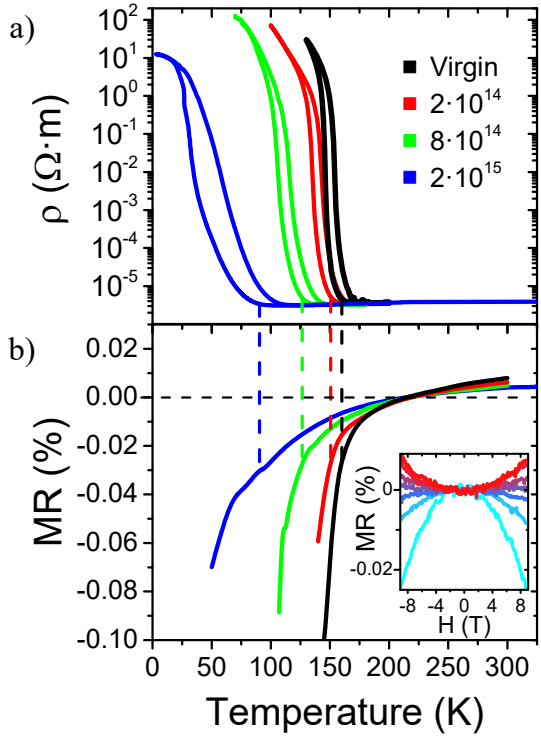

**Fig. 2 The metal-insulator transition (MIT) and its associated magnetoresistance (MR).**

Resistivity $\rho$ **(a)** and MR at 9T **(b)** as a function of temperature for a series of $V_2O_3$ samples irradiated with increasing $He^+$ doses. The doses are expressed as ions·cm$^{-2}$. The $\rho$ and MR curves are displaced together to lower temperatures as the irradiation dose is increased. The vertical dashed lines are guides to the eye to remark this fact. The MR curves show a crossover from positive to negative values, and seem to diverge as the transition takes place. Inset: Series of MR versus field $H$ curves measured at several temperatures from 280K (red) to 160K (light blue) for the virgin $V_2O_3$ sample. A quadratic dependence with field is observed.

In order to explore the relation between the MR and the MIT, we performed the same measurements for $He^+$ ion irradiated $V_2O_3$ bridges. As previously reported, ion irradiation is an ideal method to create disorder[19,20] and controllably decrease the MIT temperature in this compound[16]. We fabricated several films grown simultaneously that were irradiated with increasing doses of 100 keV $He^+$ ions: $2·10^{14}$, $8·10^{14}$, $2·10^{15}$ and $8·10^{15}$ ions·cm$^{-2}$. Fig. 2(a) shows that increased irradiation displaces the transition to lower temperatures at a rate of 4.7 K/($10^{14}$ ions·cm$^{-2}$) but without affecting the transition height (black to blue curves). Interestingly, Fig. 2(b) shows that the MR moves to lower temperatures following the shift in the $\rho(T)$ curves of Fig. 2(a). This confirms that the measured MR is associated with the MIT. Fig. 3(a),(b) show the resistivity and MR for the sample irradiated with $8 \cdot 10^{15}$ ions·cm$^{-2}$. The MIT magnitude is severely depressed (see inset), so in order to make the transition clearer we have plotted $\rho-\rho_m$ in the main panel, where $\rho_m$ is the background metallic resistivity. Importantly, due to the lower resistance of this sample, we were able to measure the MR down to the lowest temperature. Fig. 3(b) shows that the MR stops decreasing at around 50K and turns back up towards positive values.

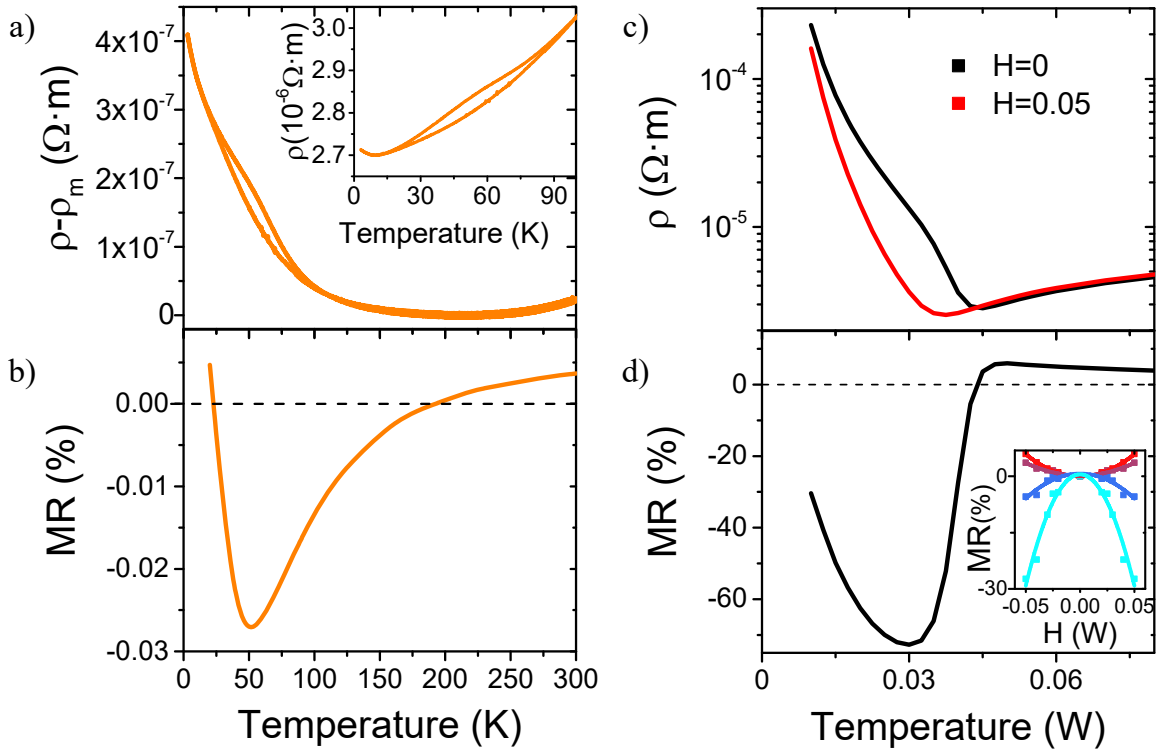

**Fig. 3 Comparison of experiments and dynamical mean field theory (DMFT) calculations.**

Resistivity $\rho$ **(a),(c)** and MR **(b),(d)** as a function of temperature. The left panels **(a),(b)** correspond to a V$_2$O$_3$ sample irradiated with $8 \cdot 10^{15}$ He$^+$ ions·cm$^{-2}$, while the right panels **(c),(d)** correspond to DMFT calculations of the Hubbard model. The inset in **(a)** shows how the MIT in this sample was weakened, but still displays the hysteresis across the resistivity upturn, a hallmark of the first order MIT. This is why in the main panel we plotted $\rho-\rho_m$, where $\rho_m = 2.3 \cdot 10^{-6} + T \cdot 7.0 \cdot 10^{-9}$ $\Omega$m is the background metallic resistivity. The black curve in **(c)** corresponds to the calculation for no applied field, and the red curve for $H$=0.05. The resistivity was calculated using the cubic root of the V$_2$O$_3$ unit cell volume as the lattice spacing. **(d)** Calculated magnetoresistance as a function of temperature. Inset: Calculated magnetoresistance (squares) for T(W)=0.05, 0.045, 0.0425, 0.04 (from red to light blue). Lines are parabola fits, showing the dependence with field is quadratic as in the experiments. Overall, the agreement between theory and experiments is remarkable.

To help gain physical insight into these results, we consider the single band Hubbard model at half filling and solve it within dynamical mean field theory (DMFT) (see methods). This basic model Hamiltonian already provided crucial insight on the paramagnetic first-order metal-insulator transition in Cr-doped $V_2O_3$[18,21,22]. Our calculations are given in units of the bandwidth $W=4t$ ($t$ is the hopping amplitude), which may be estimated around 0.8 eV according to $V_2O_3$ LDA calculations[23]. In the following, we will show how our intriguing experimental data are well reproduced and interpreted by the model calculations.

We first seek to understand the data qualitatively. For that we calculate the resistivity $\rho(T)$ for $H$=0 and $0.05W/\mu_B$ displayed in Fig. 3(c) (black and red lines respectively), where $\mu_B$ is the Bohr magneton. We observe that, upon cooling, the system undergoes a MIT in both cases but at a lower temperature for the red curve with the applied field. From the two curves, we obtain the MR as a function of $T$ shown in Fig. 3(d). Our calculations provide a good account of the anomalous features measured experimentally. Specifically, there is a small positive MR at high T followed by a decrease upon cooling that starts at $T$~0.05, just above the $T_{MIT}$ indicated by the sharp upturn of $\rho(T)$ at $T$~0.04. Also, as found experimentally, upon further lowering $T$ the (negative) enhancement of the MR becomes very steep. Eventually, it slows down until it reaches a (negative) maximum. At even lower $T$ the MR changes its behaviour and returns back towards zero. Overall, both the calculated resistivity and MR remarkably follow the experimental behaviour [see Fig. 3(a),(b) and Fig. 3(c),(d)].

We can even compare our calculations to the experimental data semi-quantitatively. Thus, we compute the field dependence of the MR in the model and find that, similarly to the experiment, *MR(H)* is proportional to $H^2$ [see inset Fig. 3(d) compared to inset Fig. 2(b)]. The quadratic dependence allows us to extrapolate to the low-field limit and estimate the *MR* at the experimental field of 9T, which yields *MR*~0.1 %, in good agreement with the measured values.

The origin of the puzzling experimental data finds a transparent physical interpretation from our model calculations. Key insight is provided by the results shown in Fig. 4. In panel (a) we plot the staggered antiferromagnetic moment $m_{AF} = -\mu_B g_S/\hbar \langle S_z \rangle$ in zero field (black line). $\langle S_z \rangle$ is the mean value of the z-component of the spin polarization of one of the antiferromagnetic sublattices and $g_S = -2$ is the gyromagnetic factor. The staggered moment signals the onset of Neel order at $T_N$~0.06. In the $H$=0 case, the magnetization in both sublattices, A and B, is equal in magnitude but opposite, $|m_A| = |m_B| = m_{AF}$ [see sketch at the bottom of Fig. 4(b)]. For non-zero $H$, this symmetry is broken. In Fig. 4(a) we plot $m_A$ (red line) and $m_B$ (blue line) for $H$=0.05W/$\mu_B$. Above $T_N$ we observe an *H*-induced polarization that is uniform in both sublattices. However, below $T_N$, the two sublattices have to oppose each other. Crucially, one of them also opposes the external field, the B sublattice in our case. Thus, near the Neel temperature, where the $m_{AF}$ is still small, sublattice B is frustrated and the moments remain disordered as they fail to polarize [see sketch at the bottom of Fig. 4(c)]. Order eventually appears at low enough $T$ when the antiferromagnetic coupling energy $\sim m^2 t^2/U$ overcomes the external magnetic field energy $\sim mH$.

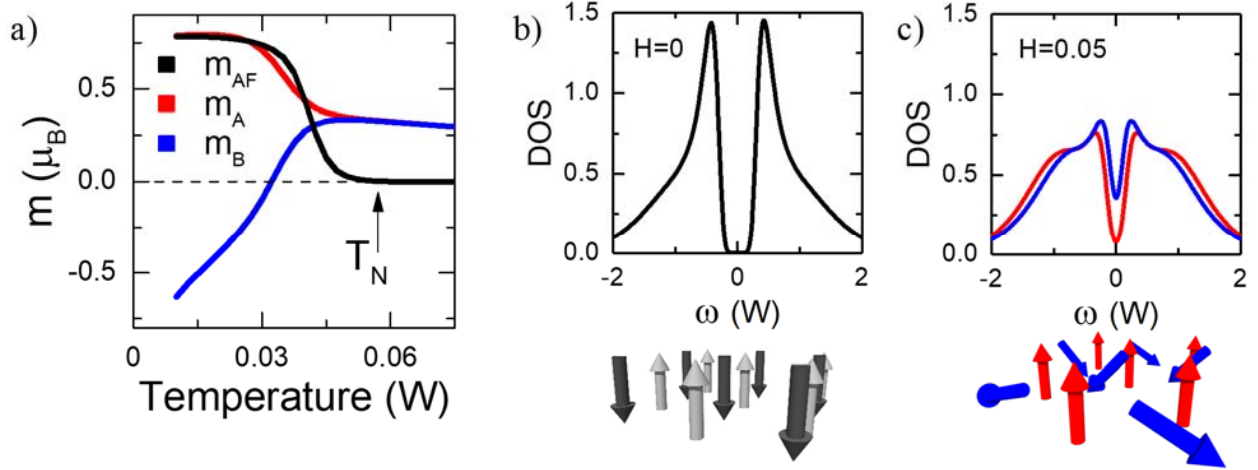

**Fig. 4 The calculated effect of the magnetic field on the antiferromagnetic (AF) ordering.**

**(a)** Calculated evolution as a function of temperature of the AF staggered moment $m_{AF}$ (black line) in zero field, and the two magnetic sublattices $m_A$ (red line) and $m_B$ (blue line) under a magnetic field $H=0.05$ parallel to sublattice A. All moments are in units of the Bohr magneton $\mu_B$ per V atom. Under a magnetic field, sublattice B does not develop its full moment due to the frustration from the magnetic field. The difference in spin ordering is depicted at the bottom of **(b),(c)**. Without an applied field, all spins (in black) order antiferromagnetically. When an external field is applied along the A sublattice (red spins), the B sublattice (blue spins) is disordered. The calculated density of states (DOS) without an applied field **(b)** and $H=0.05$ **(c)** reveals that the magnetic field produces a closing of the AF gap, more pronounced for sublattice B (blue line) than for sublattice A (red line).

Calculations of the self-energy (not shown) indicate that, with an applied magnetic field, the electronic state has a large incoherent scattering due to the failure of the B sublattice to order, which prevents the opening of a gap. This is shown in Fig. 4(b),(c). The black curve in Fig. 4(b) corresponds to the density of states (DOS) at $T=0.05$ without an applied magnetic field. A clear gap is seen at the Fermi energy, $E_F$. However, when a field is applied at the same $T$ in Fig. 4(c) the gap of sublattice B remains filled. In other words, just beneath $T_N$, $m_{AF}$ is small enough so that the external field $H$ can collapse the gap of one sublattice, the frustrated one, and gives rise to the large negative MR effect. Upon further cooling, the antiferromagnetic interactions grow as $m_B$ continues to increase until it reaches $-m_A$ at the lowest temperatures [Fig. 4(a)]. In this regime, the effect of $H$ is rendered negligible and therefore the MR returns back to zero. From this analysis, we may then argue that the observed shift of the MR curves to lower temperatures with increasing irradiation dose [Fig. 2(b)] indicates that structural disorder effectively decreases the AF coupling, thus lowering the MIT temperature. Most likely, disorder may simply renormalize down the hopping parameter $t$.

In summary, the experimental magnetotransport measurements in He-irradiated $V_2O_3$ films supplemented by Hubbard model calculations show that antiferromagnetism is the key driver of the metal-insulator transition in $V_2O_3$ upon cooling. This is indicated by (i) an unusual change of sign of the MR above the MIT, (ii) the divergent behaviour of the negative MR just beneath $T_N$, (iii) the non-monotonic behaviour at lower T, and (iv) the systematic change with the lowering of the MIT temperature. All these unusual features are captured by our model calculations and are linked to the field-induced collapse of the transport gap in one of the antiferromagnetic sublattices. The present scenario suggests that the opening of the antiferromagnetic gap (a second order transition) would subsequently provoke the structural transformation of the lattice (possibly a first order transition) as a consequence of the correlation-driven electron localization. Moreover, this may also explain the earlier intriguing neutron scattering data[17], in which the low-$T$ phase adopts a different wave-vector than the strong antiferromagnetic fluctuations found in the paramagnetic phase. This wave-vector change could also be a consequence of the electron localization and the ensuing structural change, which modify all the intersite magnetic couplings. Thus, our study reveals that the MIT in $V_2O_3$ originates in the correlation-driven opening of an antiferromagnetic gap, namely, a realization of a Mott-Slater transition.

**Methods:**

$V_2O_3$ thin films 100 nm-thick were grown epitaxially on r-cut sapphire substrates by RF magnetron sputtering deposition. A detailed account of the sample deposition and characterization can be found elsewhere[24]. Seven films grown at the same time were used for this experiment. One was left in the virgin state while the others were blanket-irradiated with increasing doses of 100 keV $He^+$ ions: $2·10^{14}$, $8·10^{14}$, $2·10^{15}$ and $8·10^{15}$ ions/cm$^2$. After irradiation 40 μm-wide and 100 μm-long microbridges for in-plane electrical transport were defined on the samples via photolithography and reactive ion etching (Ar+Cl$_2$). Ti/Au electrodes were deposited using e-beam evaporation and lift-off. All electrical transport measurements were performed in a Quantum Design Dynacool PPMS equipped with a 9T magnet. A Keithley 6221 current source was used in conjunction with a Keithley 2182A nanovoltmeter behind two Keithley 6517A electrometers acting as a buffer. This allows measuring high resistances without having the relatively low internal resistance of the nanovoltmeter in parallel with the sample.

DMFT calculations were done using a continuous-time quantum Monte Carlo impurity solver[25], based on the hybridization expansion and with a semi-circular non-interacting density of states. A one-band antiferromagnetic self-consistency with applied external magnetic field was implemented. A moderate value for the electronic interactions $U/W$=0.85 was adopted, where $U$ is the Coulomb repulsion and $W$ the bandwidth [with $W$=$4t$ and $t$ the hopping amplitude]. To avoid the inherent problems of analytical continuation of Monte Carlo data, the conductivity was obtained from the Kubo formula in bosonic Matsubara frequencies[18,26,27], and its zero-frequency value was extrapolated. To obtain the density of states, a Maximal Entropy numerical method was used. Special care was taken to obtain well converged solutions in the vicinity of the magnetic transition. For the resistivity calculation the cubic root of the $V_2O_3$ unit cell volume as the lattice spacing was used.

**Data availability.** The data supporting the plots within this paper and other findings of this study are available from the corresponding author on reasonable request.

**Acknowledgements:**

We thank Igor Mazin and Noam Bernstein for useful discussions. The magnetic measurements were supported by the Office of Basic Energy Science, U.S. Department of Energy, BES-DMS funded by the Department of Energy's Office of Basic Energy Science, DMR under grant DE FG02 87ER-45332 and the fabrication and characterization by the Vannevar Bush Faculty Fellowship program sponsored by the Basic Research Office of the Assistant Secretary of Defense for Research and Engineering and funded by the Office of Naval Research through grant N00014-15-1-2848. MR acknowledges support by public grants from the French National Research Agency (ANR) project LACUNES No ANR-13-BS04-0006-01, and the French-US Associated International Laboratory on Nanoelectronics funded by CNRS. JEV thanks support from ERC grant Nº 647100. AC gratefully acknowledges support from CONICET and UBACyT. JT and JdV would like to thank Fundación Ramón Areces for a postdoctoral fellowship.

**Authors contributions:**

This is a highly collaborative research. The experiments were conceived jointly. Samples were fabricated by JT, JdV and YK. The electronic transport measurements were done by J.T. The DMFT calculations were performed by A.C. and M.R. The data was extensively debated and the paper was written by multiple iterations between all the coauthors.

**Competing Interests:**

The authors declare no competing interests.

**Figure legends:**

**Fig. 1 Measurement device and configuration.**

Optical microscope image of the $V_2O_3$ microbridge rendered in a 3D model. For the magnetoresistance measurements the field is applied out of plane and perpendicular to the current. The voltage is measured between the longitudinal contacts.

**Fig. 2 The metal-insulator transition (MIT) and its associated magnetoresistance (MR).**

Resistivity ρ **(a)** and MR at 9T **(b)** as a function of temperature for a series of $V_2O_3$ samples irradiated with increasing $He^+$ doses. The doses are expressed as ions·$cm^{-2}$. The ρ and MR curves are displaced together to lower temperatures as the irradiation dose is increased. The vertical dashed lines are guides to the eye to remark this fact. The MR curves show a crossover from positive to negative values, and seem to diverge as the transition takes place. Inset: Series of MR versus field $H$ curves measured at several temperatures from 280K (red) to 160K (light blue) for the virgin $V_2O_3$ sample. A quadratic dependence with field is observed.

**Fig. 3 Comparison of experiments and dynamical mean field theory (DMFT) calculations.**

Resistivity ρ **(a),(c)** and MR **(b),(d)** as a function of temperature. The left panels **(a),(b)** correspond to a $V_2O_3$ sample irradiated with 8·$10^{15}$ $He^+$ ions·$cm^{-2}$, while the right panels **(c),(d)** correspond to DMFT calculations of the Hubbard model. The inset in **(a)** shows how the MIT in this sample was weakened, but still displays the hysteresis across the resistivity upturn, a hallmark of the first order MIT. This is why in the main panel we plotted ρ-$ρ_m$, where $\rho_m = 2.3 \cdot 10^{-6} + T \cdot 7.0 \cdot 10^{-9}$ Ωm is the background metallic resistivity. The black curve in **(c)** corresponds to the calculation for no applied field, and the red curve for $H$=0.05. The resistivity was calculated using the cubic root of the $V_2O_3$ unit cell volume as the lattice spacing. **(d)** Calculated magnetoresistance as a function of temperature. Inset: Calculated magnetoresistance (squares) for T(W)=0.05, 0.045, 0.0425, 0.04 (from red to light blue). Lines are parabola fits, showing the dependence with field is quadratic as in the experiments. Overall, the agreement between theory and experiments is remarkable.

**Fig. 4 The calculated effect of the magnetic field on the antiferromagnetic (AF) ordering.**

**(a)** Calculated evolution as a function of temperature of the AF staggered moment $m_{AF}$ (black line) in zero field, and the two magnetic sublattices $m_A$ (red line) and $m_B$ (blue line) under a magnetic field $H$=0.05 parallel to sublattice A. All moments are in units of the Bohr magneton $\mu_B$ per V atom. Under a magnetic field, sublattice B does not develop its full moment due to the frustration from the magnetic field. The difference in spin ordering is depicted at the bottom of **(b),(c)**. Without an applied field, all spins (in black) order antiferromagnetically. When an external field is applied along the A sublattice (red spins), the B sublattice (blue spins) is disordered. The calculated density of states (DOS) without an applied field **(b)** and $H$=0.05 **(c)** reveals that the magnetic field produces a closing of the AF gap, more pronounced for sublattice B (blue line) than for sublattice A (red line).